\documentclass[twocolumn]{aastex63}
\usepackage{verbatim}
\usepackage{comment} 

\def\be{\begin{equation}}
\def\ee{\end{equation}}

\newcommand\quotes[1]{``{#1}"}
\def\gsim{\lower.5ex\hbox{\gtsima}} 
\def\lsim{\lower.5ex\hbox{\ltsima}} 
\def\gtsima{$\; \buildrel > \over \sim \;$} 
\def\ltsima{$\; \buildrel < \over \sim \;$} \def\gsim{\lower.5ex\hbox{\gtsima}} 
\def\lsim{\lower.5ex\hbox{\ltsima}} 
\def\simgt{\lower.5ex\hbox{\gtsima}} 
\def\simlt{\lower.5ex\hbox{\ltsima}}

\def\msun{{\rm M}_{\odot}}
\def\lsun{{\rm L}_{\odot}}

\def\S*{$\Sigma_{\rm SFR}$}

\def\kms{{\rm km\,s}^{-1}\,}

\definecolor{apcolor}{HTML}{b3003b}
\definecolor{afcolor}{HTML}{800080}
\definecolor{lvcolor}{HTML}{DF7401}
\definecolor{mdcolor}{HTML}{01abdf} 
\definecolor{cbcolor}{HTML}{ff0000}
\definecolor{sccolor}{HTML}{cc5500} 
\definecolor{sgcolor}{HTML}{00cc7a}

\def\Omegap{\Omega'}

\makeatletter
\def\@hex@@Hex#1%
 {\if a#1A\else \if b#1B\else \if c#1C\else \if d#1D\else
  \if e#1E\else \if f#1F\else #1\fi\fi\fi\fi\fi\fi \@hex@Hex}
\makeatother
\definecolor{afcolor}{HTML}{b3443c}
\newcommand{\AF}[1]{({\color{afcolor} AF: #1})}

\shorttitle{Quasar feedback}
\shortauthors{Ferrara et al.}
\graphicspath{{./}{figures/}}

\begin{document}

\title{The feedback of quasars on their galactic environment}

\correspondingauthor{Andrea Ferrara}
\email{andrea.ferrara@sns.it}

\author[0000-0002-9400-7312]{Andrea Ferrara}
\affil{Scuola Normale Superiore,  Piazza dei Cavalieri 7, 50126 Pisa, Italy}

\author[0000-0003-4244-8527]{Tommaso Zana}
\affil{Scuola Normale Superiore,  Piazza dei Cavalieri 7, 50126 Pisa, Italy}

\author[0000-0002-7200-8293]{Simona Gallerani}
\affil{Scuola Normale Superiore,  Piazza dei Cavalieri 7, 50126 Pisa, Italy}

\author[0000-0002-2906-2200]{Laura Sommovigo}
\affil{Scuola Normale Superiore,  Piazza dei Cavalieri 7, 50126 Pisa, Italy}

\begin{abstract}
Quasar outflows might either quench (negative) or enhance (positive feedback) star formation in galaxies located in the quasar environment. The possible outcome depend on 4 parameters: the quasar ($\sigma$) and satellite ($\sigma_*$) halo velocity dispersion, their relative distance, $d$, and satellite disk radius, $r_d$. We find that: (i) small satellites with $\sigma _* < 164\ \sigma_{200}^{2/3}\, \rm km\ s^{-1}$ have their star formation quenched; (ii) in larger satellites, star formation, and hence UV/FIR luminosity, is instead boosted by $>80$\% in a burst with typical duration of $5-10$ Myr, if the following {positive feedback} criterion  is met: ${d}/{r_d} < 15 (Q/\eta)^{1/2} \sigma_{200}$, where $Q \approx 1$ is the satellite disk Toomre parameter; {the disruption parameter (see eq. \ref{eq:eta}) must be} $\eta>1$ to prevent complete satellite gas removal. We compare our predictions with ALMA data finding that observed satellites of $z\simeq 6$ QSOs on average form stars at a $3\times$ higher rate with respect to field galaxies at the same redshift. Further tests of the model are suggested.
\end{abstract}

\section{Introduction}
\label{sec:intro}

Amongst the most powerful and luminous ($\simgt 10^{13} \lsun$) objects in the Universe, quasars (QSOs) are thought to produce their enormous energy from the accretion of gas onto supermassive ($\gsim 10^{8} \msun$) black-holes \citep[][]{Frank_et_al_2002}. The number of known quasars has steadily increased in recent years, making them rather common sources, even at the highest redshifts so far explored \citep{Carnall_et_al_2015, Jiang_et_al_2015, Reed_et_al_2015, Venemans_et_al_2015, Wu_et_al_2015, Matsuoka_et_al_2016, Reed_et_al_2017, Banados_et_al_2018, Yang20, Wang21}.

Part of the momentum carried by radiation can be transferred to the surrounding gas of the quasar host, thereby launching powerful outflows which might extend to tens of kpc distances, often comparable to the virial radius of the host dark matter halo. Such large-scale gaseous outflows have been observed in various gas phases \citep[see, e.g.,][]{Maiolino12, Cicone2014, Carniani_et_al_2015, Fiore_et_al_2017, Fluetsch_et_al_2019, Fluetsch_et_al_2022}, and represent one of the most spectacular and effective consequences of quasar feedback \citep[][]{Silk_Rees_1998, Menci_et_al_2008, Faucher-Giguere_et_al_2012}.

Although the effect of quasar feedback onto the host galaxy has been the focus of numerous works \citep[see, e.g.,][]{Silk_Rees_1998, Bower_et_al_2006, Hopkins_et_al_2006, Barai18, Van_der_Vlugt_Costa_2019, Costa_et_al_2020}, the possibility that outflows can significantly affect galaxies residing in the quasar environment (in short, "satellites") has been, so far, only marginally considered.

On general grounds, there is no clear consensus on the effect of outflows on satellite galaxies.
At relatively low redshift, some observational and numerical studies have claimed that star formation (SF) in satellite galaxies can be \textit{quenched} because of an enhanced intergalactic medium temperature \citep[$z\sim0$;][]{Martin-Navarro_et_al_2019}, and/or gas stripping \citep[$z<3$;][]{Dashyan_et_al_2019}. Other works have suggested instead that SF can be \textit{enhanced} because of the gas density increase produced by the outflow \citep[$z\lesssim2$;][]{Croft_et_al_2006, Fragile_et_al_2017, Gilli_et_al_2019}. Additional proposed mechanisms involve the effect of quasar outflows in piercing the host gas halo, lowering its density in a bi-polar region.
This would reduce the effect of ram-pressure stripping affecting those satellites that are falling close to these regions, eventually resulting in a SF excess, with respect to the companions orbiting far from the outflows \citep[$z\sim0$;][]{Martin-Navarro_et_al_2021}.

At higher redshift, the picture is even more blurred.
Theoretical works \citep[][$2\lesssim z \lesssim 9$]{Efstathiou_1992, Thoul_Weinberg_1996, Okamoto_et_al_2008} and observations \citep[][$z\sim5$]{Kashikawa_et_al_2007} discuss the possible quenching of SF in satellites caused by QSO photoionization. \citet{Zana_et_al_2022} studied the effect of $z>6$ quasar outflows on satellites in cosmological simulations. They find that galaxies directly impacted by the outflow have their SF rate enhanced by a factor $2-3$, likely because of a gas pressure increase within the satellite.

With the aim of clarifying some of these issues, here we present a simple analytical model identifying the conditions for which QSO outflows can either suppress or enhance the SF rate of surrounding satellites.

The paper is organized as follows\footnote{Throughout the paper, we assume a flat Universe with the following cosmological parameters:  $\Omega_{\rm M}h^2 = 0.1428$, $\Omega_{\Lambda} = 1- \Omega_{\rm M}$, and $\Omega_{\rm B}h^2 = 0.02233$,  $h=0.6732$, $\sigma_8=0.8101$, where $\Omega_{\rm M}$, $\Omega_{\Lambda}$, $\Omega_{\rm B}$ are the total matter, vacuum, and baryonic densities, in units of the critical density; $h$ is the Hubble constant in units of $100\,\kms$, and $\sigma_8$ is the late-time fluctuation amplitude parameter \citep{Planck18}.}. In Sec. \ref{sec:env} we define the properties of the QSO environment in which the outflow propagates (Sec. \ref{sec:outflow}). The possible effects of the outflow shell (Sec. \ref{sec:shell}) on the satellite galaxies is discussed in Sec. \ref{sec:impact}. The main result of the work, a criterion for the outflow to produce a positive feedback (i.e., an enhancement of the SF rate in the satellites), is presented in Sec. \ref{sec:trigger}; this is then compared with available data in Sec. \ref{sec:obstest}. A summary (Sec. \ref{sec:sum}) concludes the paper.  

\section{Quasar environment}\label{sec:env}
\begin{figure}
\centering\includegraphics[scale=0.4]{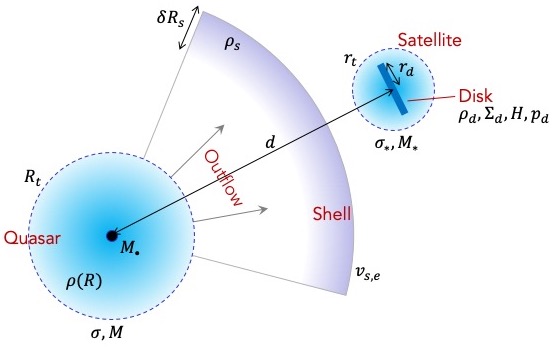}
\caption{Sketch of the model showing key symbols used.}
\label{Fig:01}
\end{figure}

Consider a quasar (QSO) hosted by a galaxy embedded in a spherical dark matter (DM) halo of total mass $M$. In addition to DM, the QSO environment is made of (a) gas, which we assume to follow the same distribution as the DM, and (b) galaxy satellites gravitationally bound to the central host galaxy. Our aim it to understand how the properties of the satellites are affected by the outflow launched by the QSO. A sketch of the scenario is shown in Fig. \ref{Fig:01},
also highlighting the key symbols used in the paper.
%
%

Assume that the gas distribution follows a singular isothermal radial density profile, 
\be
\rho_g(R) = \frac {f_g \sigma^2}{2\pi G R^2}, 
\label{eq:dens} 
\ee
where $\sigma$ is the velocity dispersion of typical QSO host galaxy spheroids, and $f_g$ is the gas fraction in the system. This number is bound from above by the cosmological value: $f_g < f_c = \Omega_{\rm B}/\Omega_{\rm M}=0.156$, but we allow for smaller values to account for the fact that an amount $M_\star$ of the baryons is contained in stars (see eq. \ref{eq:fg} below). The corresponding gas mass contained within $R$ is 
\be
M_g(<R) = \frac {2 f_g \sigma^2}{G} R. 
\label{mass} 
\ee
One can show \citep{Shapiro99} that in this case the virial temperature of the system, $T_{\rm vir} = \mu m_p \sigma^2/k_B$, where $\mu$ is the gas mean molecular weight, $m_p$ is the proton mass and $k_B$ the Boltzmann constant. Following \citet{Shapiro99}, we use this relation to define the truncation radius, $R_t$, which we take as an approximation to the virial radius, $r_{\rm vir}$, and circular velocity\footnote{We warn that the $\sigma$ value in eq. \ref{eq:sigma} is higher by $\approx 4\%$ with respect to \citet{Shapiro99} due to their assumption of a {truncated isothermal sphere} (TIS) with a central core.}, $v_c = (GM/R_t)^{1/2} \approx 1.4 \sigma$, of the system:

\begin{eqnarray}
R_t &=& 403 \left(\frac M{10^{13}M_{\odot}}\right)^{1/3}
	(1+z_c)^{-1}h^{-2/3}\,\, {\rm kpc},\label{eq:rt}\\
\sigma &=& 234\left(\frac {M}{10^{13}M_{\odot}}\right)^{1/3}
	(1+z_c)^{1/2}h^{1/3}\,\, {\rm km\,s^{-1}}.
\label{eq:sigma}
\end{eqnarray}
In the previous equations, $h$ is the nondimensional Hubble constant, and $z_c$ the halo collapse redshift, here taken to be $z_c=6$. By combining the two previous eqs. we obtain\footnote{We use the notation $Y_x = Y/x$ in appropriate units} the useful relation $\sigma_{\rm kms} = 0.58  (1+z_c)^{3/2} h\, R_{t,\rm kpc}$.

\section{Quasar outflows}\label{sec:outflow}
The QSO initially launches a radiation-pressure driven, relativistic wind ($v_w = \zeta c \simeq 0.1 c$), \citep{2003MNRAS.345..657K,2010MNRAS.402.1516K} emanating from the accretion disk powering the central supermassive black hole (SMBH). We follow \citet[][hereafter FS]{Ferrara16} and assume that the wind material has a constant velocity $v_w$ and an outflow rate $\dot M_w \equiv dM_w/dt,$ such that the momentum rate is equal to
\be
\frac{dM_w}{dt} v_w \simeq  \frac{L_E}{c},
\label{eq01} 
\ee 
where $L_E$ is the Eddington luminosity of a black hole of mass $M_\bullet$,
\be
L_E = \frac{4\pi G M_\bullet c}{\kappa_{\rm es}} = 1.54\times 10^{38} \left(\frac{M_\bullet}{M_\odot}\right) {\rm erg\, s}^{-1},
\label{eq02} 
\ee 
and $\kappa_{\rm es}=0.4 \, {\rm cm}^2$ g$^{-1}$ is the electron scattering opacity. The corresponding kinetic energy input rate is 
\be
\frac{1}{2}\frac{dM_w}{dt} v_w^2 \simeq \frac{L_E^2}{2 \dot M_w c^2} \simeq \frac{\lambda}{2} L_E \equiv L_w,
\label{eq03} 
\ee
where we have further assumed the outflow rate equals the Eddington rate, $\dot M_w \approx \dot M_E,$ and  $\lambda \equiv L_E/ (\dot M_w c^2) \simeq 0.1$ is the canonical  radiative accretion efficiency \citep{Yu02}.

The wind expansion drives a \textit{forward} spherical shock propagating in the halo, whose radius we define as $R_1$, and a \textit{reverse} shock, located at a radius $R_2 < R_1$, propagating back into the wind. In between, a contact discontinuity separates the shocked gas and wind material. 

As detailed in FS, the gas passing through the  reverse shock cools extremely rapidly due to the the strong cooling provided by inverse Compton scattering. As a result, the shocked gas rapidly loses pressure and collapses in a thin shell whose expansion is driven by momentum injection from the quasar radiation. 
Such momentum-driven phase ends quite rapidly ($\approx 10^4$ yr) as flux geometrical dilution produces a drop in the Compton cooling rate. This occurs at a radial distance from the quasar equal to      
\be
R_{C} = 0.25 \,\sigma_{200} \zeta^2_{-1}  M_{\bullet,8}^{1/2} \,\textrm{kpc},
\label{eq04} 
\ee
where $\zeta_{-1} \equiv v_w/0.1c$, and $\sigma_{200} = \sigma/200\, \kms =1$ corresponds, from eq. \ref{eq:sigma}, to a halo mass $M=6.3\times 10^{12}(1+z_c)^{-3/2} h^{-1} M_\odot$. Beyond $R_C$, the shocked wind material remains very hot and highly pressurized, so the outflowing motions are energy-driven, rather than momentum-driven.  As $R_C \ll R_t$, for our purposes we will neglect this initial phase, and consider the shell expansion in the energy-conserving regime, where the wind luminosity $L_w = (\lambda/2) L_E$ is conserved. 

According to \cite{Zubovas14} the SMBH mass at the transition to the energy-driven regime is
\be
M_\sigma =  \frac {f_g \kappa_{es} \sigma^4}{\pi G^2} = 2.3\times 10^9 f_g \sigma_{200}^4 \, M_\odot.
\label{eq05} 
\ee

Knowing the SMBH mass, we can compute $f_g$ at the onset of the outflow. Up to that point the baryonic (gas + stellar) mass is conserved: $M_g + M_\star = f_c M$. We further assume that the $M_\bullet - M_\star$ relation holds, pose $M_\bullet \simeq M_\sigma = \alpha M_\star$, and use  eqs. \ref{eq:sigma} and \ref{eq05} to obtain 
\be
f_g = \frac{f_c}{1+ (0.0045/\alpha) \sigma_{200}}.
\label{eq:fg}
\ee
Using the locally measured\footnote{The value of $\alpha$ at $z \simgt 6$ is very poorly determined.} SMBH-to-stellar mass ratio $\alpha = 0.00186$ \citep{Ding20} from the so-called $M_\bullet-\sigma$ relation, we obtain $f_g \approx 0.05$ for $\sigma_{200}=1$. For simplicity, and given the uncertainties, we will use the value $f_g=0.05$ in the following. 

The time evolution of the shell, of radius $R_s$, is found from the simultaneous solution of the momentum and energy equations \cite[][]{Weaver77}: 
\be
\frac{d}{dt}\left[M_g(< R_s) \dot R_s\right] = \frac{L_E}{c} + 4\pi R_s^2 p, 
\label{eq09} 
\ee
\be
\frac{3}{2}\frac{d}{dt}\left(\frac{4\pi}{3} R_s^3 p\right) = L_w - 4\pi R_s^2 p \dot R_s,
\label{eq10} 
\ee
where $p$ is the {(time-dependent)} pressure of the cavity bounded by the shell. We have neglected cooling losses in the energy equation as we will show (Sec. \ref{sec:shell}) that, due to the $\propto r^{-2}$ density distribution, the evolution is essentially adiabatic. We also neglect gravity as we will see below that the shell velocity largely exceeds $\sigma$. 

In the energy-driven phase, by construction, $L_E/c \ll 4\pi R_s^2 p$. By substituting the previous expressions for $M_g(R), L_w, L_E$ into eqs. \ref{eq09}-\ref{eq10}, and further assuming that $M_\bullet = M_\sigma$, we find 
\be
R_{s,e} =  \left(\frac{2\lambda c \sigma^2}{3}\right)^{1/3} t \equiv v_{s,e} t, 
\label{eq13} 
\ee
i.e. the shell moves at a constant velocity $v_{s,e} = 930 \, \sigma_{200}^{2/3} \kms$, regardless of $f_g$.

\section{Shell properties}\label{sec:shell}
A parcel of the halo gas engulfed by the forward shock, travelling at a speed $\dot R_s = {\textstyle \frac{1}{2}} (\gamma+1) {v_{s,e}} =(4/3) v_{s,e}$ for an adiabatic index $\gamma=5/3$, will be heated to a temperature
\be
T_2 =  \frac{3\mu m_p}{16 k_B} \dot R_s^2 = 2.2\times 10^7 \sigma_{200}^{4/3}  \, \textrm{K}. 
\label{eq20fs}
\ee
Assuming that the post-shock medium has a density $\rho_2=4 \rho_g(R)$, appropriate for a strong shock, and that the cooling function $\Lambda(T) = 3\times 10^{-23} (T/10^7 {\rm K})^{-0.7}$ erg cm$^3$ s$^{-1}$ \citep{Sutherland1993} in the range $10^5 \mathrm{K} \simlt T \simlt 10^7 \mathrm{K}$, we find that the cooling time is
\be
t_c = \frac{3}{8}\frac{\mu m_p k T_2}{\rho_g \Lambda(T_2)} = 0.45\, \sigma_{200}^{0.27} R_{s,\rm{kpc}}^2 \, \textrm{Myr},  
\label{eq21fs} 
\ee
where $\mu=0.65$ for the metal-enriched, ionized halo gas, and we have used eqs. \ref{eq:dens} and \ref{eq20fs}.  This timescale  is shorter than the dynamical time of the shock $t_d \simeq 3 R_s/ 4 v_{s,e} = 0.79 R_{s,\mathrm{kpc}} \sigma_{200}^{-2/3}$ Myr only for  $R_{s,\mathrm{kpc}} < 1.7\, \sigma_{200}^{-0.94} = R_{f}$.  Beyond $R_{f}$ the shock becomes adiabatic and gas cannot cool in a thin shell. Nevertheless, most of the swept-up mass piles up in a \textit{thick} shell of hot gas located behind $R_s$. Moreover, as $R_{f} \ll R_t$ we can safely neglect the brief initial shock radiative phase resulting in a thin shell. The structure function, i.e. the distribution of the mass within the cavity carved by a non-radiative wind, depends on various parameters and ambient gas density distribution (for a full calculation see e.g. the classic papers by \citet{Weaver77, Truelove99}). 

Here a more approximate treatment will suffice for our scopes. To compute the thickness of the shell, $\delta R_s$ we simply use mass conservation of the halo swept-up gas ending up in the shell:
\be
M_g(<R_s) = \frac {2 f_g \sigma^2}{G} R_s =  4\pi \rho_2(R_s) R_s^2  \delta R_s,
\ee
which gives $\delta R_s = R_s/4$.

As the shock and the thick shell run over a satellite galaxy orbiting the halo, two outcomes are in principle possible: (a) the ISM of the galaxy might be stripped by the ram pressure of the impinging outflow, or (b) SF can be boosted by disk gas compression. We analyse these two scenarios separately in the following Sections.

\section{Satellite gas removal}\label{sec:impact}
We start by considering the condition leading to (a). Early works studied the problem of a gas cloud of radius $R_{cl}$ run over by a shock, and the resulting instabilities. Basically, the shock may induce Kelvin-Helmoltz (KH) instabilities that strip the gas from the cloud. However, these authors showed that the instability can be suppressed if the cloud is bound by a sufficiently strong gravitational potential. For example, \citet{Murray93, Vietri97} found that the stability of a gravitating cloud against disruption by a shock travelling at velocity $\dot R_s$ is guaranteed when the critical parameter,
\be
\eta = \frac{g \Delta R_{cl}}{2\pi \dot R_s^2} > 1, 
\label{eq:eta}
\ee
where $\Delta$ is the cloud overdensity with respect to the background medium, and $g$ is the cloud surface gravity. The above condition can be recasted in a cosmological framework \citep[][see also \citealt{Sigward05}]{Cen08} in which the cloud is a satellite galaxy embedded in a halo with properties similar to the central galaxy, i.e. a truncated isothermal sphere with velocity dispersion $\sigma_*$, see Sec. \ref{sec:env}. 

In this situation, $g = G M(<r)/r^2 = 2 \sigma_*^2/r$, where $r$ is the satellite galactocentric radius, $\Delta = 18 \pi^2$ is the halo overdensity according to the nonlinear collapse theory, and $\dot R_s = (4/3) v_{s,e}$. We evaluate $\eta$ as a function of $r$, thereby substituting $R_{cl}$ with $r$, but it turns out that $\eta$ is independent of $r$: 
\be
\eta = \frac{81}{8} \pi \left(\frac{\sigma_*}{v_{s,e}}\right)^2. 
\ee
The stability condition, $\eta > 1$, sets the minimum value of $\sigma_*$ for the satellite to survive the shock passage: 
\be
\sigma_* > \sqrt{\frac{8}{81\pi}} v_{s,e} = 0.177\ v_{s,e} = 164\ \sigma_{200}^{2/3}\, \rm km\ s^{-1}
\label{eq:ablate}
\ee
From the previous equation, we see that only relatively massive satellites can retain their gas, while smaller ones have their gas stripped by the outflow. As a rule of thumb, eq. \ref{eq:ablate} states that satellites with halo mass $\simlt 1/2$ of the QSO host one have their gas removed by the outflow.

\section{Enhanced star formation}\label{sec:trigger}
We now analyse the second possibility (b). In satellites satisfying eq. \ref{eq:ablate}, the outflow travels at a velocity $v_{s,e}$ which is larger than the escape speed from the satellite. However, the gas in the shell is decelerated by a shock as it enters the satellite halo, cools\footnote{We speculate cooling of the shocked gas at the outflow-satellite interface should produce copious Ly$\alpha$ emission, as recently observed by \citet{Vito20}.}, and eventually joins the ISM of the galaxy.  

As the shell impacts on the satellite galaxy disk, it enhances its pressure. The ISM gets compressed, cools and gets denser thereby stimulating a burst of SF\footnote{A similar mechanism has been proposed by \citet[][]{Zubovas13} to explain SF bursts in the QSO host galaxy.}. The outflow exerts a pressure, $p_s$,  given by
\be
p_s = \frac{1}{2} \rho_s v_{s,e}^2 = \frac{f_g \sigma^2}{\pi G}\frac{1}{d^2} v_{s,e}^2
\label{eq:shock_pres}
\ee
where the gas density in the shell is 
\be
\rho_s = \frac{2 f_g \sigma^2}{\pi G}\frac{1}{d^2} = 2\times 10^{-23} \left(\frac{\sigma_{200}}{d_{\rm kpc}}\right)^2 \,\, {\rm g\,cm}^{-3},
\ee
and we have used the relation $\rho_{s}=4 \rho_g$; $d$ is the radial distance of the satellite from the QSO at the impact time.
Numerically we obtain 
\be
p_s = 8.65\times 10^{-8} d_{{\rm kpc}}^{-2} \sigma_{200}^{10/3} \quad {\rm erg\, cm}^{-3}.
\ee

We now model the satellite galaxy as a disk also located in a isothermal halo as in eq. \ref{eq:dens} with 
velocity dispersion $\sigma_*$. The disk is assumed to be in centrifugal equilibrium, and rotate at a (Keplerian) angular frequency $\Omega = \sqrt{2} \sigma_*/r$. 
It follows that the disk Toomre parameter $Q$ is 
\be
Q = \frac{\kappa_\Omega c_s}{\pi G \Sigma_d} = \frac{\sqrt{2} \Omega c_s}{\pi G \rho_d H},
\label{eq:Qdef}
\ee
where $c_s^2=p_d/\rho_d$ is the gas isothermal sound speed (taken to be $c_s=10\ \kms$); $\rho_d$, $\Sigma_d$, $H$, $p_d$ are the disk volume and surface density, scale height, and pressure, respectively. We have also introduced the epicyclic frequency $\kappa_\Omega^2 = (2\Omega/r)d(\Omega r^2)/dr = 2 \Omega^2$ for a Keplerian disk ($\kappa_\Omega^2=4\Omega^2$ for a flat rotation curve). In the following we will assume that the disk is marginally Toomre-stable, $Q\approx 1$, as observed in many spiral and starburst disks, also at high redshift \citep[][and references therein]{Krumholz15}. 

For a thin disk in a spherical potential, the equation for vertical (i.e. along the $\xi$ coordinate) hydrostatic equilibrium, $\partial p_d/\partial \xi = \rho_d \nabla \phi$. Using the above definition of $\Omega$, an approximate solution for the disk pressure is then 
\be
p_d \approx \rho_d \Omega^2 H^2.
\label{eq:pres}
\ee
Thus, it is $H = c_s/\Omega$.

When the shell impacts the disk, it adds an external pressure equal to $p_s$. For simplicity, we assume that the disk is oriented face-on with respect to the shell propagation direction (angle $\theta=0$); this maximises the effect of the outflow on the satellite\footnote{The predictions for a generic inclination angle $\theta$ can be obtained by substituting $p_s$ with $p_s\cos \theta$ in eq. \ref{eq:shock_pres}.}. Such compression is physically equivalent to an enhanced effective gravitational field: by reducing the scale height, it induces a density enhancement. 

If the outflow pressure compresses the disk without ablating a significant amount of gas, which is largely preserved by the gravitational pull of the dark matter halos (see Sec. \ref{sec:impact}), then $\Sigma_d = \rho_d H \approx$ const. during the passage of the shell.
Moreover, if the cooling time of the disk gas is short, which is the case given the high densities, the transformation is close to isothermal ($c_s \approx$ const.). These two facts entail that $Q \propto c_s/\Sigma_d$ remains $\approx 1$ through the compression.  

We then rewrite eq. \ref{eq:pres} as
\be
p \approx \rho_d \Omega^2 H^2 + p_s,
\ee
and, by rearranging the terms, obtain
\be
H = \frac{c_s}{\Omega}\left[ 1 - \frac{1}{2} {\cal M}^2 \frac{\rho_s}{\rho_d}\right]^{1/2} \equiv \frac{c_s}{\Omegap},
\label{eq:H}
\ee
where ${\cal M}=v_{s,e}/c_s$ is the Mach number of the shell with respect to the disk sound speed. The last term defines the new (and higher) angular frequency $\Omegap$ that embeds the effects of the outflow pressure on the disk. The final step is to derive an expression for $\rho_d$ by substituting eq. \ref{eq:H} into eq. \ref{eq:Qdef}, 
\be
\rho_d = \frac{\sqrt{2}}{\pi}\frac{\Omega \Omegap}{G Q}
\ee
to obtain the relation 
\be
\Omegap^2 - \left(\frac{\Omega_s^2}{\Omega}\right) \Omegap - \Omega^2 = 0
\ee
which expresses $\Omegap$ as a function of $\Omega$. We have introduced the frequency $\Omega_s^2(\sigma, d) = (Q/2\sqrt{2}){\cal M}^2 \pi G\rho_s \propto t_{\rm ff}^{-2}$ which is associated with the free-fall time, $t_{\rm ff}$, of the shell gas. 
Numerically,
\be
\Omega_s = 1.16\times 10^{-13} Q\, \sigma_{200}^{5/3}\,d_{{\rm kpc}}^{-1} \quad \rm s^{-1}.
\label{eq:omegas}
\ee
The solution is 
\be
\Omegap = \frac{\sqrt{\Omega_s^4+4\Omega^4} + \Omega_s^2}{2\Omega}
\ee
Note that it is always $\Omegap > \Omega$, and as $\Omega_s\to 0$,  $\Omegap \approx \Omega$.
%
%
\begin{figure}
\centering\includegraphics[scale=0.4]{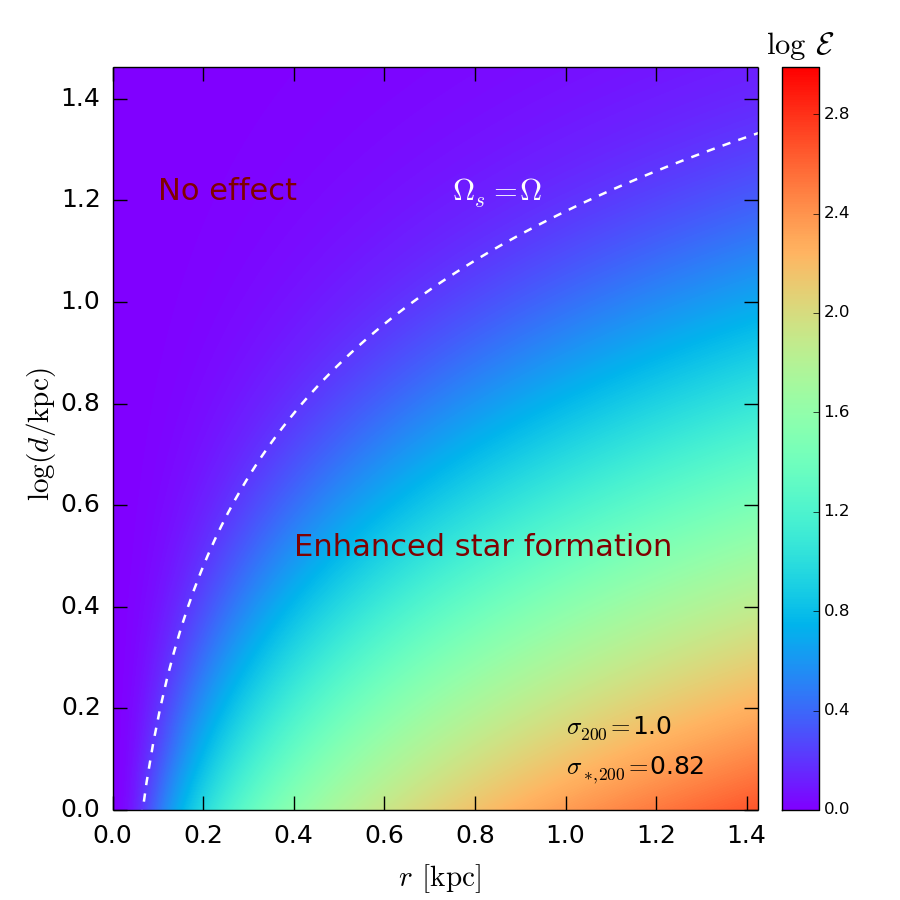}
\caption{Outflow-induced enhancement, ${\cal E}$, of the star formation in a QSO satellite galaxy with velocity dispersion $\sigma_{*,200}=0.82$ corresponding to the marginal value $\eta=1$ for complete gas removal given by eq. \ref{eq:ablate}; smaller satellites are destroyed by hydrodynamic instabilities. The enhancement value (see colorbar) is shown as a function of the satellite galactocentric radius, $r$, and its distance from the QSO, $d$; see sketch in Fig. \ref{Fig:01}. Also shown for reference is the line $\Omega_s = \Omega$, where ${\cal E}=1.62$, defining the positive feedback criterion eq. \ref{eq:criterion}.}
\label{Fig:02}
\end{figure}
The surface SF rate is usually taken to occur, both in normal and starburst galaxies, at a fraction $\epsilon \approx 0.02$ \citep{Kennicutt98, Krumholz15} of the local dynamical timescale\footnote{In the following we assume that $\Sigma_d(r) \approx$ const. with radius $r$. This is approximately true for disk galaxies, at least in the more strongly star forming central parts \citep{Wang14}}:
\be
\dot\Sigma_{\star}(r) = \epsilon\, \Omega(r)\, \Sigma_d
\label{eq:sfr}
\ee
During the passage of the outflow stars form at a rate given by $\Omegap$ rather than by $\Omega$. The SF is then enhanced by the ratio
\be
{\cal E} =\frac{\Omega'}{\Omega} = \frac{\sqrt{\Omega_s^4+4\Omega^4} + \Omega_s^2}{2\Omega^2} > 1.
\label{eq:enh}
\ee
Eq. \ref{eq:enh} is graphically depicted in Fig. \ref{Fig:02}, where we have set $\sigma_{200}=1$ and $\sigma_{*,200}=0.82$, corresponding to $\eta=1$, i.e. a satellite marginally surviving KH instability disruption according to eq. \ref{eq:ablate}. SF is enhanced
(${\cal E}>1$) in a large region of the $(r, d)$ parameter space. The SF increase is mild at the satellite center, but it increases considerably at large galactocentric radius, reaching at $r= 1$ kpc values as high as ${\cal E} \simeq 100$ up to distances $d \simeq 4$ kpc from the QSO. The enhancement clearly decreases with $d$, but it is still 1.62 (as defined by the line $\Omega_s = \Omega$) at 16 kpc.

By integrating eq. \ref{eq:sfr} over the galactocentric radius, and using eq. \ref{eq:enh}, we find that the total galaxy luminosity, $L \propto \dot\Sigma_{\star}$ (in general this is true for both the UV and IR luminosity, see e.g. \citealt[][]{Madau14}) integrated over the disk radius, $r_d$, is enhanced by the outflow by a factor (see App. \ref{app:A} for a full derivation) 
\be
\frac{L'}{L} \simeq \frac{1}{6} \left(\frac{\Omega_s r_d}{\sigma_*}\right)^2.
\label{eq:luminosity}
\ee
Thus, for the case shown in Fig. \ref{Fig:02}, a satellite (QSO) located at $d=10$ kpc, with $\sigma_{*,200}=0.82$ ($\sigma_{200}=1$), and taking $\Omega_s$ from eq. \ref{eq:omegas}, the luminosity increases by 80\% within $r_d=1$ kpc (and by 7 times for $r_d=3$ kpc). 

\subsection{Positive feedback criterion}
As a simple criterion to estimate whether the QSO feedback can trigger a SF burst in a companion galaxy we require that $\Omega_s > \Omega$; from eq. \ref{eq:enh} this condition entails  ${\cal E} =\Omega'/\Omega > (1+\sqrt{5})/2=1.62$. 

Starting from the definitions of $\Omega_s$ and $\Omega$ in Sec. \ref{sec:trigger}, the condition can be written as 
\be
\sqrt{\frac{Qf_g}{2\sqrt{2}}}\left(\frac{\sigma}{c_s}\right) \left(\frac{r}{d}\right)  \left(\frac{v_{s,e}}{\sigma_*}\right) >1.
\label{eq:semi-final}
\ee
Then, recalling the expression for $\eta$ in eq. \ref{eq:ablate}, one can substitute for the last factor in the previous equation, finally obtaining a physical criterion for positive feedback:
\be
\frac{d}{r} < 15 \sqrt{Q}\, \frac{\sigma_{200}}{\eta^{1/2}},
\label{eq:criterion}
\ee
which can be evaluated, for example, at the disk radius $r=r_d$. The criterion relates the quenching (by gas removal) parameter $\eta$ (eq. \ref{eq:ablate}) and the ratio $d/r_d$ between the satellite distance from the QSO and its disk radius for a given value of $\sigma_{200}$ of the QSO host galaxy. 

\begin{figure}
\centering\includegraphics[scale=0.38]{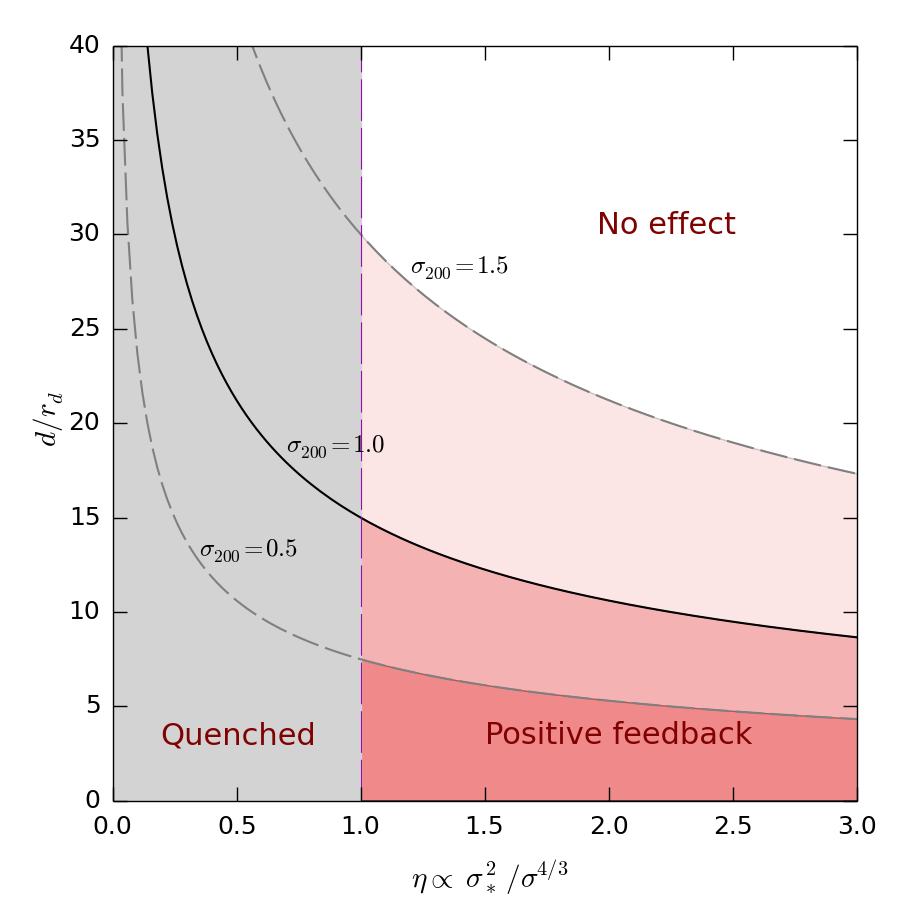}
\caption{Possible fates for a satellite engulfed by a QSO outflow as a function of the quenching parameter $\eta$ (eq. \ref{eq:ablate}) and the ratio $d/r_d$ between the satellite distance from the QSO and its disk radius (see Fig. \ref{Fig:01}). In addition to the fiducial value $\sigma_{200}=1$ (solid line), the (dashed) curves for $\sigma_{200}=0.5,1.5$ are also shown. SF in small satellites ($\eta < 1$, grey area) is quenched; larger ones have their SF rate enhanced (positive feedback, red area) if they are located sufficiently close to the QSO when they are hit by the outflow. }
\label{Fig:03}
\end{figure}

Three possible outcomes of the outflow-satellite interaction are possible, as depicted in Fig. \ref{Fig:03}. 
If $\eta < 1$, implying satellite halo masses $\simlt 1/2$ of the QSO host, the satellite gas is wiped away by KH instability and its SF suppressed. If the satellite survives the passage of the shock, the fate depends on the location of the satellite at that moment. For $\sigma_{200}=1$, satellites located within $(10-15)\, r_d$ from the QSO have their SF/luminosity enhanced by a factor $> 1.62$ (the precise value can be computed from eq. \ref{eq:luminosity}). At larger distances, the satellite is basically unaffected by the outflow. Clearly, less/more luminous quasars residing in less/more massive halos (as an example we show in Fig. \ref{Fig:03} the cases $\sigma_{200}=0.5,1.5$) extend their positive feedback to smaller/larger distances.

The duration of the burst is approximately equal to the time during which the disk gas gets compressed and then re-expands to the original configuration after the shock passage: $\delta t_s = \delta t_c + \delta t_e$. The compression time can be simply computed as $\delta t_c=\delta R_s/v_{s,e} = d/4v_{s,e} = 0.26\, \sigma_{200}^{-2/3} d_{\rm kpc}$ Myr. The re-expansion time is equal to $\delta t_e= H/c_s = \Omega^{-1} = r_d/\sqrt{2}\sigma_*$. Hence, the total duration of the burst is   
\be
\delta t_s = 0.26\, \sigma_{200}^{-2/3} d_{\rm kpc} + 3.5\,   \sigma_{*,200}^{-1} r_{\rm d,kpc}\, \rm Myr
\label{eq:duration}
\ee
Hence, for a configuration in which $\sigma_{200}=1, d_{\rm kpc}=10, \sigma_{*,200}=0.82$ (corresponding to $\eta=1$), and $r_{\rm d,kpc}=1$,
we find that the duration of the burst $\approx 7$ Myr.
%
%

\section{Observational tests}\label{sec:obstest}
Our model predicts that QSO outflows can quench, enhance or leave star formation unchanged in satellite galaxies, depending on the conditions set by eqs. \ref{eq:ablate} and \ref{eq:criterion}, and graphically shown in Fig. \ref{Fig:03}. These conditions depend on four parameters ($\sigma, \sigma_*, d, r_d$) which can in principle be obtained from observations of QSO--satellite systems. 
However, at present, the available data, which at high-$z$ are essentially restricted to those analyzed in \citet[][]{venemans:2020} and \citet[][]{Trakhtenbrot:2017}, do to not allow to derive such parameters in a reliable way. Let us discuss the critical issues. 

The values of $\sigma$ (and, analogously, $\sigma_*$) can be obtained from the measured [CII] line FWHM. We recall that for a truncated isothermal sphere $\sigma=v_c/1.4$, and $v_c= 0.75\, {\rm FWHM}/\sin\,i$, where $i$ is the galaxy inclination \citep[][]{Ho07,Venemans15}. The inclination angle is usually derived from the observed minor-to-major axis ratio: $i = \cos^{-1}(a_{\rm min}/a_{\rm max})$, assuming a thin disk geometry. 

Unfortunately, due to the limited spatial resolution of the experiment, \citet[][]{venemans:2020} provide the axis ratio only for the 27 QSO hosts in their sample, but not for the (27) detected satellites. This prevents us to obtain a precise value of $\sigma$ and $\sigma_*$ from the [CII] line FWHM. Nevertheless, by making the very rough assumption that $\sin\,i=0.7$ (the mean over the QSO sample) for both the satellites and the QSO hosts, we find $\langle \eta \rangle = 1.25\pm 0.53$. Although very uncertain, this value suggests that indeed the observed satellite galaxy population overall fulfils the condition $\eta > 1$, thus having survived the potential gas removal by the outflow. 

A much more refined analysis is impossible with the available data. This problem is even more severe when trying to apply the positive feedback criterion (eq. \ref{eq:criterion}) which requires the knowledge of $d$ and $r_d$. The former quantity suffers by the uncertainty in the determination of the satellite position along the line of sight (the projected separation is instead reasonably determined), which is heavily affected by the unknown peculiar velocity of the system. The measurement of the satellite disk radius, $r_d$, would instead require higher spatial resolution observations. Future experimental efforts will likely overcome both these problems.   

\begin{figure}
\centering\includegraphics[scale=0.38]{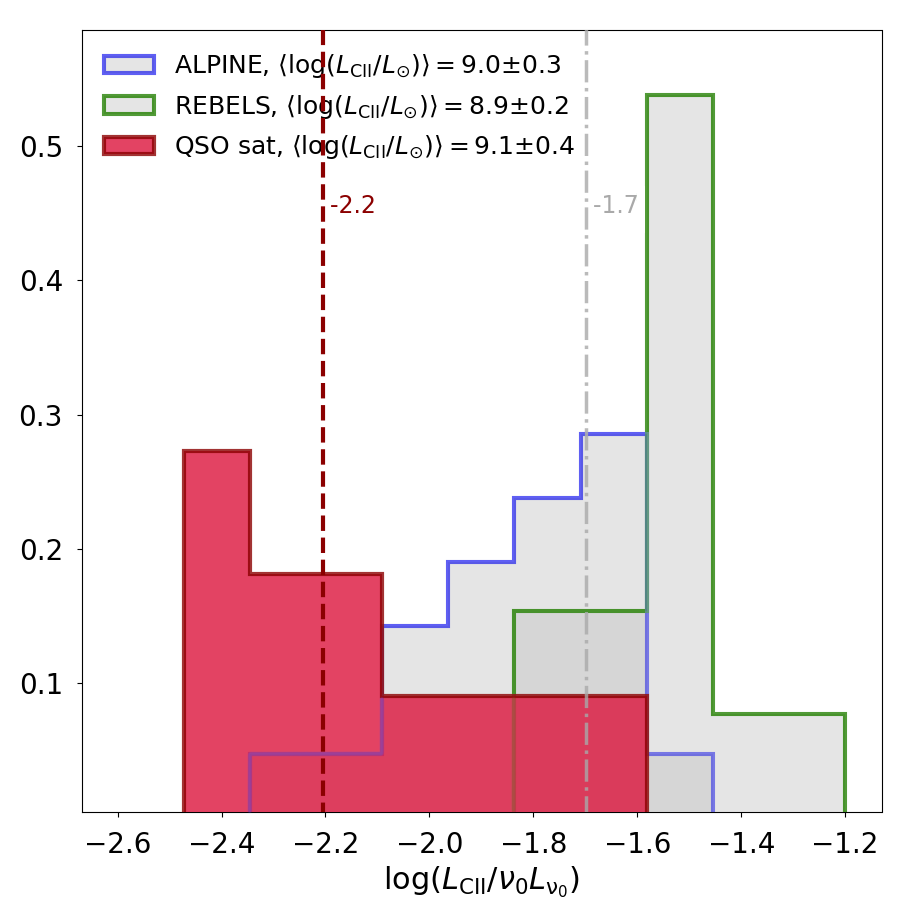}
\caption{Measured ratio between the [CII] luminosity, $L_{\rm CII}$, and the FIR continuum monochromatic luminosity, $\nu_0L_{\nu_0}$, at rest-frame $\nu_0=1900\ \mathrm{GHz}$ (wavelength $\lambda_0=158\ \mathrm{\mu m}$) in galaxies at $z>5$. We show the distribution for REBELS (green, $z\sim 7$, \citealt{bouwens:2021}), ALPINE (blue, $z\sim 5$, \citealt{lefevre:2020}), and quasar satellite (red, $z\sim 4.5-6$, \citealt{Trakhtenbrot:2017,venemans:2020})  galaxies. The median $L_{\rm CII}/L_{\rm 1900}$ value for REBELS+ALPINE galaxies, and QSO satellites is shown by a vertical dashed line (in grey and red, respectively). Despite all these galaxies have comparable $L_{\rm CII}$ (see legend), the QSO satellites show a lower $L_{\rm CII}/L_{\rm 1900}$ ratio with respect to field galaxies, implying a larger FIR luminosity, and hence SF rate.}
\label{Fig:04}
\end{figure}

In spite of the above difficulties, we can still perform some useful, albeit preliminary, comparison with available data. To this aim, from the  \citet[][]{venemans:2020} and \citet[][]{Trakhtenbrot:2017} data we build the distribution of the ratio, $L_{\rm CII}/\nu_0L_{\nu_0}$, between the [CII] and FIR continuum luminosity at rest-frame frequency $\nu_0 = 1900\ \mathrm{GHz}$, for $z \simgt 5$ quasar satellites. The  distribution spans the (log) range from $-2.5$ to $-1.6$, with a median value of $-2.2$. In Fig. \ref{Fig:04} such distribution is compared with that obtained for $z \simgt 5$ field galaxies by two ALMA surveys, i.e. ALPINE and REBELS. We note a striking (2.5$\sigma$) difference between the two distributions. Field galaxies have on average a 3 times lower [CII]/continuum ratio. Given that the satellite and field galaxy sample have an essentially identical mean $L_{\rm CII} \approx 10^9 L_\odot$, the difference can only be due to a higher FIR continuum luminosity of the satellites. As the FIR luminosity is proportional to the SF rate, we conclude that SF in satellites is generally enhanced, indicating a positive feedback effect. 

Let us interpret this evidence in the framework of our model, which predicts that SF is enhanced when the condition $\eta >1$ \textit{and} the positive feedback criterion are simultaneously satisfied. Galaxies that are quenched by the outflow ($\eta < 1$) have low mass. Hence, not only they are intrinsically faint, but their luminosity is further decreased by SF quenching. As a result, observations are biased toward massive satellites which are made even brighter by positive feedback. 

\section{Summary and discussion}\label{sec:sum}
We have presented a model for the interaction between a QSO outflow and its satellite galaxies with the aim of determining whether the outflow affects the star formation (SF) in these systems. A priori, the outflow might (i) strip the gas from the satellite by inducing powerful instabilities, thus quenching SF (negative feedback), or (ii) compress the gas leading to an enhancement of the SF activity (positive feedback). We have clarified the possible outcomes of the interaction depending on the system configuration parameters, such as the QSO ($\sigma$), and satellite ($\sigma_*$), halo velocity dispersion, their relative distance, $d$, and satellite disk radius, $r_d$. We find that:
\begin{itemize}
\item[{\color{red} $\blacksquare$}] Small satellites with a velocity dispersion $\sigma_* < 164\ \sigma_{200}^{2/3}\, \rm km\ s^{-1}$ have their SF quenched by instabilities and lose the majority of their gas. 
\item[{\color{red} $\blacksquare$}] SF, and hence UV/IR luminosity, in larger satellites is  typically boosted by 80\%; even more sizeable boosts are possible under the condition expressed by eq. \ref{eq:luminosity}. The SF burst typically lasts $5-10$ Myr.
\item[{\color{red} $\blacksquare$}] We have derived a simple criterion (eq. \ref{eq:criterion} leading to a SF enhancement ${\cal E} > 1.62$:
\be
\frac{d}{r_d} < 15 \sqrt{Q}\, \frac{\sigma_{200}}{\eta^{1/2}}
\ee
where $Q$ is the satellite disk Toomre parameter; $\eta>1$ is required to prevent complete gas removal from the satellite. Such \quotes{positive feedback} criterion is graphically shown in Fig. \ref{Fig:03}. 
\item[{\color{red} $\blacksquare$}] We have tested the model against recent ALMA data by \citet[][]{venemans:2020}. We show that, in agreement with model predictions, observed satellites of $z\simeq 6$ QSOs have on average  (i) $\langle \eta \rangle =1.25 \pm 0.53$, indicating that these systems were able to survive the passage of the outflow, (ii) a $3\times$ higher SF rate due to the outflow positive feedback.
\end{itemize}

Before concluding a few points are worth emphasising. When comparing the model results with data we have assumed that the radial distance $d$ at the time at which the outflows overruns the satellite corresponds to the observed one. This does not account for non-circular orbits. We have also assumed that the satellite disk is oriented perpendicular to the outflow velocity vector, and that the satellite velocity can be neglected compared to the outflow one. While these can be considered as reasonable assumptions not affecting the general validity of the conclusions, more detailed work, probably requiring dedicated numerical simulations, is needed to assess the actual impact of these effects. 

Another simplification regards the imposed a spherical outflow geometry. If instead the outflow is, e.g. biconical, some of the satellites will not be overrun by the shell, and therefore they will not be affected by the processes described here. Progress on this issue can only come from experimental constraints on the outflow geometry. Such model refinements would be rather straightforward, and could provide a better statistics of the feedback effects on the satellite population of a given QSO. 
Finally, one should consider the possibility that the QSO luminosity is intermittent, i.e. the observed value significantly different from its long-term average. This aspect has been preliminarly considered by \citet[][and references therein]{Zubovas20}, but its role in the present context needs to be further examined.

\acknowledgments
We thank R. Gilli for useful comments. AF acknowledges support from the ERC Advanced Grant INTERSTELLAR H2020/740120. Generous support from the Carl Friedrich von Siemens-Forschungspreis der Alexander von Humboldt-Stiftung Research Award is kindly acknowledged (AF). 
Plots in this paper produced with the \textsc{matplotlib} \citep{Hunter07} package for \textsc{PYTHON}.    

{\bf Data Availability}\\
Data available on request.

\appendix
\section{Luminosity enhancement}\label{app:A}
We provide here the derivation of eq. \ref{eq:luminosity} quantifying the enhancement of the satellite luminosity due to the (positive) feedback outflow. 
We assume that the total galaxy luminosity is $L \propto \dot\Sigma_{\star}$, which generally holds for both the UV and FIR bands \citep[][]{Madau14}. Then, before the arrival of the outflow it is
\be
L \propto 2\pi \int_0^{r_d} \dot\Sigma_{\star} r dr = 2\pi \int_0^{r_d} \epsilon \Omega(r) \Sigma_g r dr = 2\pi \epsilon \Sigma_g \int_0^{r_d} \sqrt{2} \frac{\sigma_*}{r}\dot\Sigma_{\star} r dr = 2\sqrt{2}\pi \epsilon\Sigma_g\sigma_*r_d
\label{eq:Lderiv}
\ee
We can write and analogous expression for the enhanced luminosity, $L'$, resulting form the gas compression induced by the outflow by simply substituting $\Omega$ with $\Omega'$ in the second equality of the previous equation: 
\be
L' = 2\pi \int_0^{r_d} \epsilon \Omega'(r) \Sigma_g r dr = 2\pi \epsilon \Sigma_g \int_0^{r_d} \Big[\frac{\sqrt{\Omega_s^4+4\Omega^4} + \Omega_s^2}{2\Omega}\sqrt{2}\Big] r dr.
\label{eq:L'deriv}
\ee
after some straightforward manipulation of eq. \ref{eq:L'deriv} we obtain
\be
L' = \frac{2\pi \epsilon \Sigma_g\Omega_s^2}{2\sqrt{2}\sigma_*} r_d^3 \Big[\int_0^1 dy \sqrt{y^4+ a^4} + \frac{1}{3}\Big], 
\label{eq:L'deriv2}
\ee
where $a = (2\sigma_*/\Omega_s r_d)$. The integral can be written in terms of the incomplete hypergeometric function, and it is equal to $a^2\, _2F_1(-1/2, 1/4; 5/4;-1/a^4$). We note that for most situations of interest here, $a \ll 1$. In that case, the integral is $\approx 1/3$. By taking the ratio $L'/L$ we get to the final expression in eq. \ref{eq:luminosity}  
\be
\frac{L'}{L} \simeq \frac{1}{6} \left(\frac{\Omega_s r_d}{\sigma_*}\right)^2.
\ee

\bibliography{paper}{}
\bibliographystyle{aasjournal}



\end{document}